\documentclass{emulateapj}
\usepackage{apjfonts}

\newcommand\simless{{\thinspace \rlap{\raise 0.5ex\hbox{$\scriptstyle  {<}$}}
    {\lower 0.3ex\hbox{$\scriptstyle  {\sim}$}} \thinspace }}  
\newcommand\simgreat{{\thinspace \rlap{\raise 0.5ex\hbox{$\scriptstyle  {>}$}}
    {\lower 0.3ex\hbox{$\scriptstyle  {\sim}$}} \thinspace }}  
\newcommand\msun{\, \rm M_\odot} 
\newcommand\rsun{\, \rm R_\odot}
\newcommand\kms{{\, \rm km\,s^{-1}}}
\newcommand\gyr{\, \rm Gyr} 
\newcommand\kpc{\, \rm kpc}
\newcommand\cmq{{\, \rm cm^{-2}}}
\newcommand\xte{\rm XTE~J1118+480}
\newcommand\mbh{M_{\rm bh}}

\lefthead{Gualandris et al.}
\righthead{An asymmetric natal kick in XTE J1118+480?}

\begin{document}

\title{Has the black hole in XTE J1118+480 experienced an asymmetric natal kick?}

\author{Alessia Gualandris\altaffilmark{1},
Monica Colpi\altaffilmark{2}, Simon Portegies Zwart \altaffilmark{1} 
and Andrea Possenti\altaffilmark{3}}
\altaffiltext{1}{University of Amsterdam, Kruislaan 403, 1098 SJ, 
  Amsterdam, the Netherlands}
\altaffiltext{2}{Universit\`a degli Studi di Milano Bicocca, 
Piazza della Scienza 3, 20100 Milano, Italy}
\altaffiltext{3}{INAF-Osservatorio Astronomico di Cagliari, 
loc. Poggio dei Pini, Strada 54, 09012, Capoterra, Italy}

\begin{abstract}
  We explore the origin of the Galactic high latitude black hole X-ray
  binary $\xte$, and in particular its birth location and the
  magnitude of the kick received by the black hole upon formation in the
  supernova explosion. Our analysis is constrained by the evolutionary
  state of the companion star, the observed limits on the orbital
  inclination, the Galactic position and the peculiar velocity of the
  binary system.  We constrain the age of the companion to the black
  hole using stellar evolution calculations between 2\,Gyr and 5\,Gyr,
  making an origin in a globular cluster unlikely. 
  We therefore argue that the system was born in the Galactic disk, 
  in which case the supernova must have propelled it in its current 
  high latitude orbit.
  Given the current estimates on its position in the sky, 
  proper motion and radial velocity, we back-trace the orbit of $\xte$ in the
  Galactic potential to infer the peculiar velocity of the system 
  at different disk crossings over the last $5 \gyr$. 
  Taking into account the uncertainties on the velocity components, we
  infer that the peculiar velocity required to change from a Galactic
  disk orbit to the currently observed orbit is $183\pm31\kms$.  The
  maximum velocity which the binary can acquire by symmetric supernova
  mass loss is about $100\kms$, which is 2.7\,$\sigma$ away from the
  mean of the peculiar velocity distribution. We therefore argue that
  an additional asymmetric kick velocity is required.
  By considering the orientation of the system relative to the plane
  of the sky, we derive a 95\% probability for a non null component of 
  the kick perpendicular to the orbital plane of the binary. 
  The distribution of perpendicular velocities is skewed to lower 
  velocities with an average of $93^{+55}_{-60}\kms$.
  These estimates are independent of the age of the
  system, but depend quite sensitively on the kinematic parameters of
  the system. A better constraint on the asymmetric kick velocity
  requires an order of magnitude improvement in the measurement of the
  current space velocity of the system.
\end{abstract}

\keywords{stars: black-holes -- stars: individual (XTE J1118+480) -- stars: binary -- X-ray}

\section{Introduction}

The velocity dispersion of the low mass X-ray binaries hosting  a
black hole (Liu, van Paradijs \& van den Heuvel 2001) is around 
$40 \kms$, as derived by White \& van Paradijs (1996) from a study of 
the vertical distribution of these systems in the Galaxy.
Since the velocity dispersion of the population of their progenitors 
(i.e. binary systems containing a star more massive than 
$\sim 20 \msun$, Portegies Zwart, Verbunt, Ergma 1997) 
is estimated to be $\sim 20 \kms$ (Mihalas \& Binney 1981), 
there is evidence for the acquisition of an extra velocity during 
the formation of the black hole.  There are two possible mechanisms 
to accelerate a binary system when a massive star forms and
explodes as a supernova. 
The first is the ejection of mass from the binary (Blaauw 1961). 
The released mass continues to move with the velocity of its center of mass 
and, for conservation of linear momentum, the binary recoils in the 
opposite direction. We will refer to this velocity acquired by the binary 
as {\it symmetric kick}. The direction of the symmetric kick lies in 
the orbital plane of the binary which means that the direction of 
the orbital angular momentum vector of the system is conserved by 
the explosion.  The second mechanism is known as {\it asymmetric kick} 
and represents an additional kick generated by asymmetries seeded
in the collapsing core that are transmitted to the ejecta.
These natal kicks can have random directions relative to the orbital plane. 
There is evidence for the occurrence of asymmetric kicks in the
formation of neutron stars (see Lai, Chernoff \& Cordes 2001 for a review)
but the physical interpretation is still controversial: 
mechanisms proposed include large scale density asymmetries in the 
pre-supernova core (Burrows \& Hayes 1996), or non-axisymmetric 
instabilities in the rapidly rotating proto-neutron star core 
(Colpi \& Wasserman 2002).
These hydro-dynamically driven impulses can in principle operate
at the time of black hole formation.  This is expected in particular if  
a hot proto-neutron star forms first, launching  a successful 
shock wave, and later collapses to a black hole due to falling back
material (Heger et al. 2000). The highest neutron star kick recorded is
of $800-1600 \kms$ (Cordes, Romani \& Lundgren 1993) 
so if kicks scale approximately with the inverse of the mass of the 
compact object, peculiar velocities as large as $100 \kms$ are 
expected for black holes. 
The first discovered runaway black hole is in the binary X-Ray Nova 
GRO J1655-40 (Mirabel et al. 2002) for which there is evidence of 
a motion of $112 \pm 18 \kms$. 
GRO J1655-40 is a source that shows alpha-elements in its optical
spectrum indicating that a supernova explosion preceded the formation 
of the black hole (Israelian et al. 1999).

In this work we focus on $\xte$, a soft X-ray transient 
(i.e. a low-mass X-ray binary exhibiting recurrent bright optical 
and X-ray outbursts, often accompanied by radio activity, 
Tanaka \& Shibazaki 1996, Campana et al. 1998) with a mass function 
large enough ($f(M)\approx 6.0 \pm 0.4 \msun,$ McClintock et al. 2001) 
to ensure it hosts a black hole.
Three features distinguish $\xte$ in the sample of 14 similar systems
(Lee, Brown \& Wijers, 2002; Orosz et al. 2002): {\it (i)} the
high-galactic latitude ($l=157.7^{\circ}$, $b=62.3^{\circ},$
corresponding to a distance of $1.9 \pm 0.4\kpc$ from the Sun with
a height of $1.7 \pm 0.4\kpc$ above the Galactic plane, Wagner et
al. 2001); {\it (ii)} the high space velocity ($\sim 145 \kms$ relative to
the Local Standard of Rest, Mirabel et al. 2001, much larger than the
aforementioned velocity dispersion of the population); {\it (iii)} the
shortest orbital period $P = 4.08\, \rm hours$ (Cook et al. 2000). 
Optical photometry (Wagner et al. 2001) suggests that the donor  
is a $\sim 0.3 \pm 0.2\msun$ star now filling its Roche lobe  
of $\sim 0.35 \rsun$.  Modeling of the light curve (McClintock et al. 2001)
indicates that the inclination of the system is high, $i \simgreat
55^{\circ}$, and the mass of the black hole is consequently modest,
$\mbh \simless 10\msun$.  Additional evidence for a high
inclination ($i \simgreat 60^{\circ}$) comes from measurements of tidal
distortion (Frontera et al. 2001) whereas the lack of dips or eclipses
for a Roche-lobe filling secondary yields upper limits of 
$i < 80^{\circ}$ and $\mbh \simgreat 7.1 \msun$.  The extremely low 
hydrogen column depth ($N_{\rm H} \approx 1.3 \times 10^{20}\cmq$) 
and the modest distance make this system a primary target
for the study of the origin and the dynamics of Galactic black holes.

Mirabel et al. (2001) noticed that the orbit of $\xte$ resembles
that of halo objects like Galactic globular clusters. 
In addition, the radial and azimuthal components of the velocity 
seem consistent with the large random motions of old halo stars with
low metallicities. According to these characteristics, 
Mirabel et al. suggested that the black hole in $\xte$ may be one of 
the black holes ejected from globular clusters which are believed to
swirl around in the halo of the Galaxy.  
This hypothesis requires the binary system to be coeval with the 
globular clusters.

In this paper we exploit recent results (Haswell et al. 2002) on
the evolution of $\xte$, which hint for a much younger age of the system.
We show that an ejection from a globular cluster is unlikely (\S~2) and
propose that the system originated in the disk of the Galaxy and was
then launched at high latitude by a large kick induced by the
supernova explosion. In \S~3 we investigate the nature of this
kick,  by back-tracking the orbit of $\xte$ in the Galactic potential
in order to derive  the distribution of the runaway velocity and of its
component perpendicular to the orbital plane, signature 
of an asymmetric explosion in the formation of the black hole.
In \S~4 we summarize our conclusions.

\section{The birthplace of the system}

Spectroscopic studies of $\xte$ provide important clues
about the nature of the system and the evolution of its
orbital parameters.  The UV spectrum of $\xte$ (Haswell et al. 2002) 
show evidence of an under-abundance of carbon compared to
nitrogen which suggests that the material accreted by the black hole has
been substantially CNO-processed.  This implies that the companion
star has lost its outer layers and is now exposing inner regions 
which have been enriched with CNO-processed material from the central core. 
Haswell et al. interpret these qualitative observations as an indication 
that the system came into contact when the companion star was sufficiently 
massive to allow for the CNO cycle and was already somehow evolved. 
These requirements constrain the orbital period at contact $P_i$ 
to be longer than about 12 hours and the companion mass to be 
about $1.5 \msun$.  The evolution of the system after contact 
was driven by angular momentum losses, mainly magnetic braking, 
toward shorter orbital periods.
The estimate of the initial mass of the companion star is crucial to
this investigation as it allows a determination of an upper limit for
the age of the system. The latter can in turn constrain the system's
birthplace to be the Galactic disk or a globular cluster.

Supernova explosions of massive stars occur in the first $10^7$ years
since the formation of a globular cluster, producing neutron
stars and black holes. The age of $\xte$ is hence
determined by the main sequence lifetime of the companion
star and by the binary evolution phase after contact.
A $1.5 \msun$ star has a main sequence lifetime $T_{\rm MS} \sim 2 \gyr$. 
The duration of the mass transfer phase can be estimated by means of 
binary evolution calculations.  Using the code by Eggleton (Pols et al. 1995), 
J.Dewi (private communication) followed the evolution of a system 
composed by a black hole of $7 \msun$ and a main sequence companion 
of $1.4\msun$ with an initial period $P_i = 15$ hours. 
The binary shrinks, reaching the period ($P = 4.08$ hours) currently
observed in $\xte$  after a time $T_{\rm ev} \sim 0.8 \gyr$ 
(Fig. 1 reports the evolution of the mass of the companion star as 
a function of the binary orbital period). An upper limit for the 
age of the system can then be determined as
\begin{equation}
\tau = T_{\rm MS} + T_{\rm ev} \sim 3 \gyr.
\end{equation}
If, according to the uncertainties in the evolutionary models,
we allow for a range of initial masses of the companion star $1.2
\msun \simless {\it m} \simless 1.8 \msun$, the main sequence lifetime for
the companion is limited to the interval $1 \gyr \simless
T_{\rm MS} \simless 4.5 \gyr,$ and, as a consequence, the age
of the system must lie in the range $2 \gyr \simless \tau
\simless 5.5 \gyr$.  
On the basis of these estimates we can
exclude that the system was formed and then ejected from 
one of the known Galactic globular clusters, which halted forming
stars more than $10 \gyr$ ago
\footnote{There exists a possibility that the black hole in $\xte$
  experienced a dynamical encounter in a globular cluster
  with a blue straggler in a binary system tight enough
  to lead to the ejection of the black hole and the newly
  acquired companion. Stellar encounters involving black
  holes are effective in the first billion year since the cluster
  formation (Portegies Zwart \& McMillan 2000) as the black holes tend 
  to interact with each other, form binaries which harden through 
  subsequent interactions and eject each other from the cluster
  (Sigurdsson \& Hernquist 1993; Kulkarni, Hut, McMillan 1993). 
  This scenario requires that at least one black hole is retained 
  and that blue stragglers are continuously formed in the core, 
  preferentially in a hard binary. We consider this possibility very unlikely.}.

There exist further uncertainties associated with the evolution of $\xte$
which may make somehow affect the estimate of $\tau$.
In particular, differences in the modeling of magnetic braking can 
introduce a scatter in the value of $\tau$: a weaker braking would 
cause the binary to harden up to the semi-detached state on a longer time. 
Other uncertainties are related to the observed properties
of the companion star, like the luminosity, effective temperature and 
metallicity, which may affect the estimate of the age of the system.
A quantitative evaluation of such uncertainties would require
detailed studies which are beyond the aim of this work. 
Nonetheless, conciliating $\tau$ with the hypothesis of the origin of the
binary in a globular cluster would demand an unlikely fine-tuning of the
binary evolution. 
In this paper we assume a standard model for the binary evolution,
like the one adopted by Haswell et al. (2002), according to which
$\xte$ is too young to have been ejected from a globular cluster \footnote{
The only scenario compatible with an origin in a cluster is that of a black
hole ejected $\sim 10 \gyr$ ago from a cluster and then capable of
capturing a main sequence star of about $1.5 \msun$ while
wandering through the Galaxy: the probability for such an event in the
low stellar density of the Galaxy is in fact negligible.}  
and hence we conclude that the system was very likely born in the disk of the
Galaxy.

\begin{figure}[!ht]
\includegraphics[width=7.5cm]{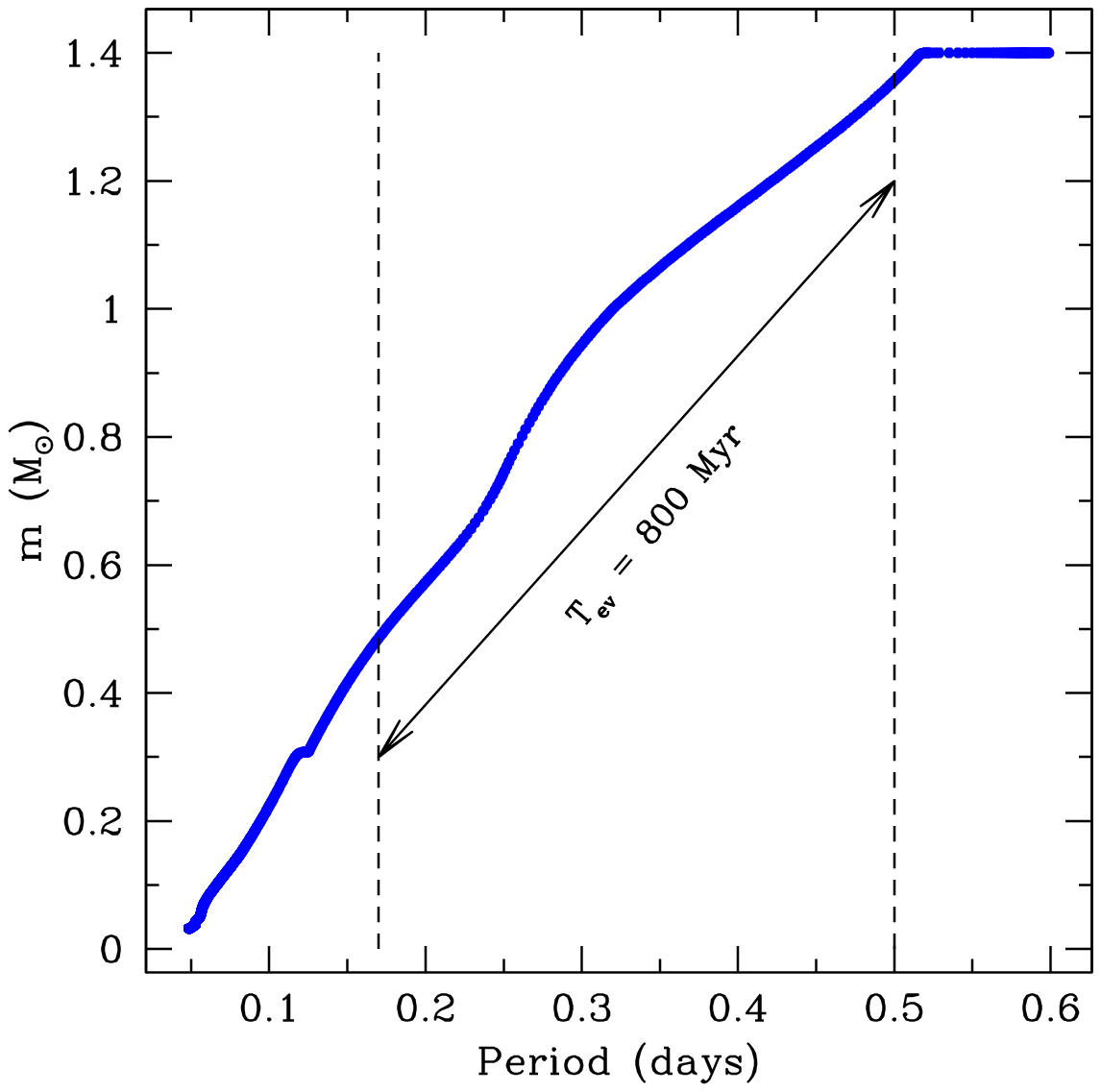}
\caption{Evolution of a binary system composed 
  by a black hole of $7 \msun$ and a main sequence star of 
  $1.4 \msun$ with an orbital period at first contact 
  $P_i = 0.5$ days. Angular momentum losses drive the shrinking 
  of the binary until, after a time $T_{\rm ev} \simeq 800 \,\rm Myr$,
  the system reaches the current period of about 0.17 days. 
  At that point the companion star has a mass of about 
  $0.48 \msun$, compatible with the observed mass 
  $m = (0.3 \pm 0.2) \msun$ of the donor in $\xte$.}
\end{figure}

\section{The supernova kick hypothesis: 
  back-tracking the orbit of XTE J1118+480 in the Galactic potential}

We now consider the hypothesis that the system originated in the
disk of the Galaxy and was launched at high galactic latitude by
a kick induced by the supernova explosion which produced the
black hole.  Given the large mass of the black hole progenitor, its
evolution must have been very fast, whence we can assume that the
supernova event took place shortly after the beginning of the main
sequence life of the donor star. However, the time of expulsion 
$\tau_{\rm ex}$ from the Galactic disk is affected by the 
uncertainties in the duration of the X-ray phase 
and in the mass interval of the donor star. 
The length of the X-ray phase determines a lower limit of $0.5-1 \gyr$ 
since mass transfer necessarily begins after the black
hole formation. 
On the other hand, an upper limit of $2-5 \gyr$ reflects 
the age of the companion star on the main sequence. 
More specifically, the lower and upper values refer to a star which is now 
respectively at the beginning or at the end of its main sequence phase.
Combining these limits, for the most probable mass of the donor star $m = 1.5 \msun$ 
(Haswell et al. 2002) we obtain $0.8 \simless \tau_{\rm ex} \simless 3 \gyr$.

\subsection{Integration of the galactic orbit of XTE J1118+480 backward in time}

The integration of the trajectory of $\xte$ backward in time
provides the value of the kick velocity acquired at birth
by the black hole when residing in the Galactic disk.
It is our aim to compute this value which is nonetheless affected 
by a number of uncertainties. The orbit of the system is in fact not unique 
given the finite interval of expulsion times 
and the observational errors in the velocity vector.
The observed proper motion and the radial velocity are known with 20\% and 50\%
accuracy, respectively.  Thus, a statistical approach is necessary
for the study of the orbit.  
We generate the initial conditions for the integration by drawing at random the 
three components of the velocity vector from Gaussian distributions. 
Once we have initialized the velocity components for 10000 different Monte Carlo 
realizations, we trace the trajectories backward in time 
with a fifth order Runge-Kutta integrator using 
the Paczynski model (Paczynski 1990) for the potential of the Galaxy. 

In Figure 2 we show the position of the source in the Galactic plane 
obtained integrating up to a time $0.5\gyr$ (left panel) and, 
continuing the integration, up to $5\gyr$ (right panel). 
We notice that, due to the uncertainties in the current 3D velocity,
memory is lost of the position of the binary at the time of formation,
if the explosion occurred more than $1 \gyr$. 
(We verified that a change of 10\% in the parameters of the Galactic
potential results in differences of at most 5\% in the velocity
of the system.)

\begin{figure*}[!ht]
\begin{center}
\includegraphics[width=7cm]{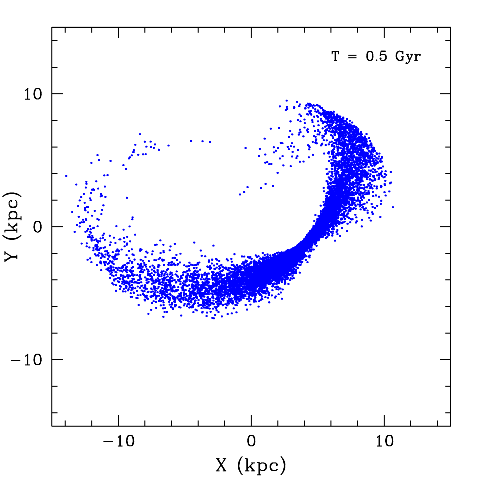}
\includegraphics[width=7cm]{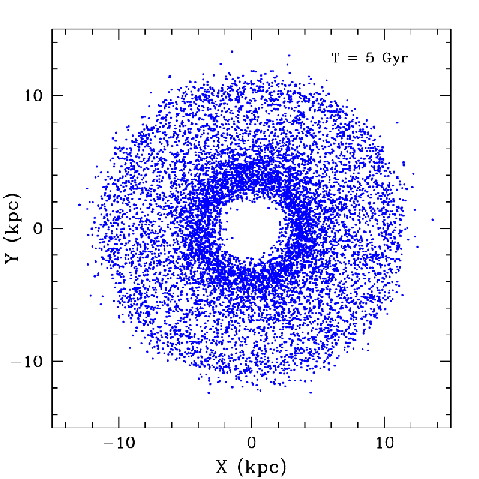}
\caption{Position of $\xte$ in the Galactic plane (the Sun being at a position
  X=-8.5 kpc, Y=0) at an epoch
  of $0.5\gyr$ ago (left plot) and $5\gyr$ ago (right plot) 
  obtained from 10000 integrations of its trajectory backward in time 
  starting from its current position 
  ($l = 157.7^{\circ}$, $\delta = +62.3^{\circ}$, $d = 1.85\kpc$). 
  The velocity components are randomly drawn from Gaussian distributions 
  ($U=-105\pm16 \kms$, $V=+122\pm16 \kms$, $W=-21\pm10 \kms$, 
  relative to the Galactic center) using the Monte Carlo code described in \S~3.1.
  The uncertainties on the current 3D velocity of the system propagate
  throughout the integration with a resulting loss of information
  on the position of the system at any time in the past anterior 
  to about $1 \gyr$ ago. However, the shape of the orbit is such 
  that the system always remains at a distance greater than about $2\kpc$
  from the Galactic center.}
\end{center}
\end{figure*}

Figure 3 shows the distributions of the modulus of the peculiar
velocity of the system (after correcting for rotation around the 
Galactic center) obtained from the Monte Carlo realizations corresponding 
to the passages through the disk closer to some specified time. 
In particular, we select the crossings closest to 0.5, 2, 3 and 5 Gyr 
in the past. The distributions for the peculiar velocity tend to 
widen while increasing time backward, but the mean and the shape 
of the distributions do not change considerably at the 
different passages through the disk, allowing to safely adopt as 
reference value for the peculiar velocity the average computed over 
all the disk crossings between 0.5 and 5 Gyr ago, 
$<V_p> = 183 \pm 31 \kms$.
\begin{figure}[!ht]
\begin{center}
\includegraphics[width=8.2cm]{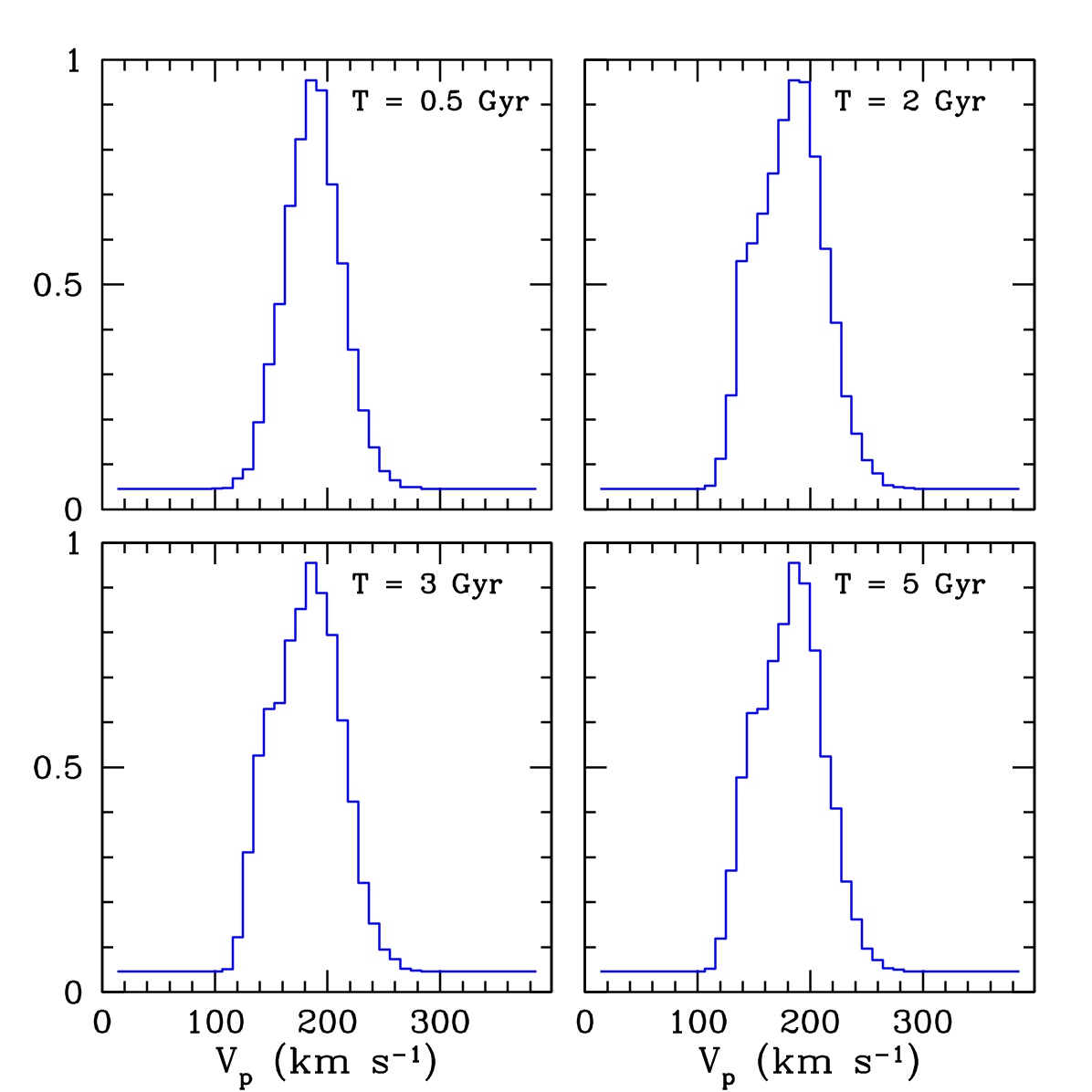}
\caption{Probability distributions for the peculiar velocity
   of $\xte$ at different disk crossings as obtained from the 
   integration of 10000 orbits backward in time, where the current 
   velocity components are drawn randomly from Gaussian distributions 
   centered around the observed values. The distributions 
   at different disk crossings do not differ significantly. 
   This property allows the determination of an average mean peculiar 
   velocity at the moment of the ejection from the Galactic disk 
   which is independent of the exact time of the ejection.}
\end{center}
\end{figure}

We first explore (\S~3.2) the possibility that a symmetric kick 
is entirely responsible for the large peculiar velocity of $\xte$
and show that this scenario is very unlikely. 
In \S~3.3 we combine the available information on the 3D 
orientation of the binary in the celestial sphere and the results
on its dynamics through the Galaxy to put constraints 
on the contribution from an additional asymmetric kick, 
intrinsic to the supernova event.

\subsection{Symmetric kick}

The amount of mass $\Delta M$ which has to be ejected during a
symmetric supernova explosion in order to produce a kick velocity
$V_{\rm sym}$ can be written as a function of the black hole mass
$\mbh$ and the companion mass $m$ (Bhattacharya \& van den Heuvel,
1991; Nelemans, Tauris, van den Heuvel 1999) as
\begin{equation}
\left(\frac{\Delta M}{\msun}\right) = \left(\frac{V_{\rm sym}}{213
\kms}\right) \left(\frac{P}{\rm day}\right)^{\frac{1}{3}}
\left(\frac{\mbh + m}{\msun}\right)^{\frac{5}{3}}
\left(\frac{\msun}{m}\right)
\end{equation}
where $P$ is the orbital period of the circularized binary after the
explosion.  The system remains bound after the supernova only if the
amount of ejected mass is smaller than half the total mass of the
system before the explosion, $\Delta M < 0.5 (m + M_{\rm He})$, where
we indicate with $M_{\rm He}$ the mass of the black hole progenitor.
If we replace $M_{\rm He} = \mbh + \Delta M$, the above condition can
be written as $\Delta M < (\mbh + m)~$.

The peculiar velocity of the system at the most recent disk crossing,
as determined by Mirabel et al. (2001), is $V_p = 217 \pm 18 \kms$.
They used this value of the velocity together with the current mass 
($\sim 0.3 \msun$) of the companion star and the current orbital 
period to estimate the ejected mass needed in a symmetric supernova 
explosion. They found that more than $40 \msun$ should have been 
ejected during the stellar collapse in order to accelerate
the black hole system up to the velocity $V_p$. This value is
implausibly large for the binary to remain bound.

In order to improve on that calculation, we have integrated the orbit
of the system backward in time in the Galactic potential and derived 
the peculiar velocity of the binary center of mass at different disk
crossings. As shown in \S~3.1, the average peculiar velocity
of the binary computed over all the disk crossings between
0.5 and 5 Gyr ago is $<V_p> = 183 \pm 31 \kms$.
We now explore the possibility that $\xte$ acquired its high
peculiar velocity as a result of a symmetric kick.  
The maximum symmetric kick compatible with the survival of the binary 
as a function of the orbital period after circularization is shown in Fig. 4. 
It is obtained when the mass ejected in the explosion is maximum, 
$\Delta M = \mbh + m~$.
The three regions refer to values of the donor star mass $m$ of 1.2,
1.5 and $1.8\msun$, respectively, from bottom to top.
For each value of $m$, the black hole mass vary in the range $6.0-8.0\msun$.
The orbital period of the circularized binary is allowed to vary
between a minimum value, corresponding to the case where 
the companion star fills its Roche Lobe and the system comes into
contact after recircularization, and a maximum value set by the constraint 
that the donor star fills its Roche lobe within its lifetime because 
of magnetic braking (Kalogera 1999).
The maximum kick velocity obtainable with a symmetric supernova
explosion is then $V_{\rm sym, max} \simeq 120 \kms$, not enough to
impart a recoil to $\xte$ as large as the observed peculiar velocity. 
In fact, the distribution of the peculiar velocity is such that there is 
a very low probability (0.9\%) for $V_p \le 120 \kms$ at a disk crossing
(see Fig. 3).
In particular, for the reference case ($m=1.5\msun$, $\mbh=7\msun$), 
$V_{\rm sym, max} \simeq 100 \kms$.
\begin{figure}
\begin{center}
\includegraphics[width=7.5cm]{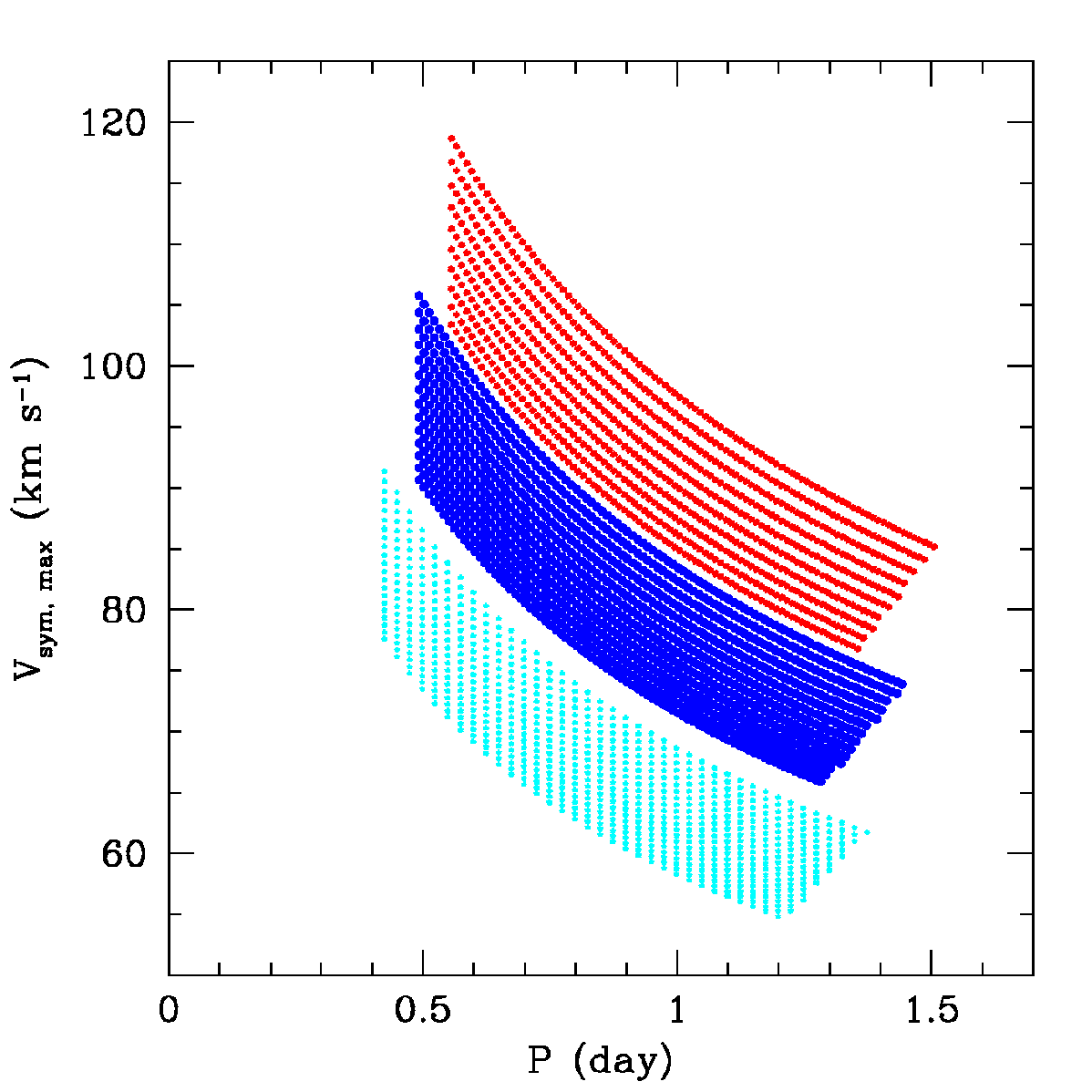}
\caption{ Maximum recoil velocity imparted to the binary center of
  mass in a symmetric supernova explosion as a function of the
  orbital period after the recircularization phase.
  The minimum value for the orbital period is set by the condition
  that the companion star fills its Roche lobe after
  recircularization while the maximum value is set by the constraint 
  that the donor star fills its Roche lobe within its lifetime 
  because of magnetic braking.
  The three regions refer to values of the donor star mass $m$ of 1.2,
  1.5 and $1.8\msun$, respectively, from bottom to top.
  For each value of $m$, the black hole mass vary in the range
  $6.0-8.0\msun$.}
\end{center}
\end{figure}

\subsection{Asymmetric kick}

The considerations of the \S~3.2 indicate that an asymmetric
kick is likely to be needed in addition to a symmetric kick to produce
the large space velocity of $\xte$.  A lower limit on the value of the
asymmetric kick imparted to the system can be obtained if the
direction of the orbital angular momentum vector of the system
is known.  Estimates of the inclination of the orbital plane of the
binary relative to the line of sight are available from optical
spectroscopy and photometry combined with theoretical models,
giving $60 ^{\circ} \simless i \simless 80 ^{\circ}$ (Wagner et
al. 2001, Frontera et al. 2001).  We will consider an
intermediate $i = 70^{\circ}$ as a reference value.  Unfortunately,
the position angle $\Omega$ of the line of nodes with respect to the
celestial coordinates is not known and therefore the exact
3D-orientation of the orbital plane of the binary can not be
reconstructed. However, the components of the angular momentum vector 
can be written as a function of $i$ and $\Omega.$ If we
randomly draw values for the undetermined angle from a uniform
distribution in the range $[0 - 2\pi)$, we can compute -- at each disk
crossing -- the component $V_{p,\perp}$ of the peculiar velocity of
$\xte$ which is perpendicular to its orbital plane.  Since the
symmetric kick lies in the orbital plane while the asymmetric kick can
have any direction, the component $V_{p,\perp}$ provides a lower limit
for the magnitude of the asymmetric kick, once corrected for the
contribution of the random motion of the binary's progenitor
(about $10 \kms$ for one component).

Fig. 5 reports the distribution for $V_{p,\perp}$
at disk crossing from the 10000 Monte Carlo realizations which are
closer to some specified times in the past, namely 0.5 Gyr (the
likely lower limit for $\tau_{\rm ex}$) and 2, 3, 5 Gyr ago,
corresponding to the upper limit of $\tau_{\rm ex}$ for a companion
star with initial mass of 1.8, 1.5 and 1.2 $\msun$ respectively.  
These histograms have broad peaks covering a range in $V_{p,\perp}$ 
from 60 to 140 $\kms$. They show that the hypothesis of a null 
intrinsic kick in the direction perpendicular to the orbital plane has 
a probability of at most 5\%. Such probability corresponds to all 
the Monte Carlo realizations of the motion of $\xte$ in the Galaxy 
for which $V_{p,\perp}$ at disk crossing is less than the unidimensional 
dispersion velocity of the progenitor of the binary system.
\begin{figure}
\begin{center}
\includegraphics[width=8.2cm]{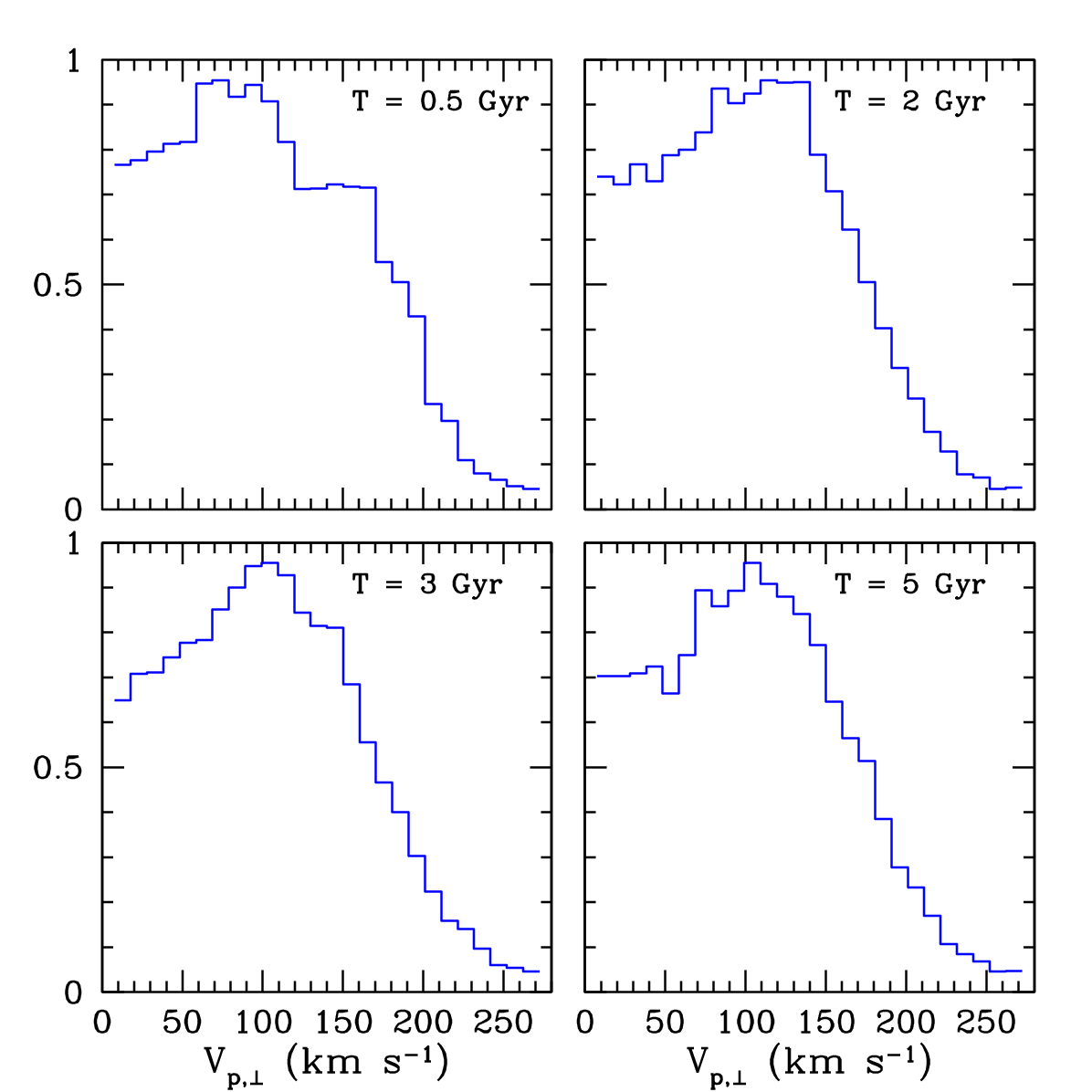}
\caption{Probability distributions for the component of the peculiar
   velocity of $\xte$ which is perpendicular to its orbital
   plane. The histograms are obtained with the orbit integrator
   described in \S~3.1 assuming an inclination angle of
   the orbital plane $i = 70^{\circ}$. Distributions at different disk
   crossings, between 0.5 and 5 Gyr ago, are represented.}
\end{center}
\end{figure}

The distributions of $V_{p,\perp}$ presented in Fig. 5 are wider 
and flatter than those of $V_{p}$ (Fig. 3) since they result from 
the overlap of the distributions obtained for any possible value 
of $\Omega$ (in the upper panel of Fig. 6 we display a set
of these distributions, for the case $\tau_{\rm ex}=3 \gyr$ ago). 
The dominant source of indetermination arises from the uncertainties on
the current kinematic parameters of $\xte$ and on the
time of occurrence of the supernova event $\tau_{ex}.$ 
As an example, comparison of the upper and central panels of Fig. 6 
shows that even the knowledge of $\Omega$ with a
$10^\circ$ accuracy could not ensure a significant determination
of $V_{p,\perp}$, unless the errors on the measured radial and
transverse velocity of $\xte$ were reduced down to about 1\%. 
In particular, the large uncertainties on the radial velocity
($V_r = 26 \pm 17 \kms$, McClintock et al. 2001; $15 \pm 10 \kms$, 
Wagner et al. 2001) 
sensibly affect the global error on the system's space velocity.
The lower panel in Fig. 6 displays the effect of the
indetermination on $\tau_{ex}$: the distributions of $V_{p,\perp}$
summed over all the disk passages between 0.5 and 3 Gyr ago
have broader peaks than those corresponding to the single passage 
at about 3 Gyr ago.
\begin{figure}
\begin{center}
\includegraphics[width=8.2cm]{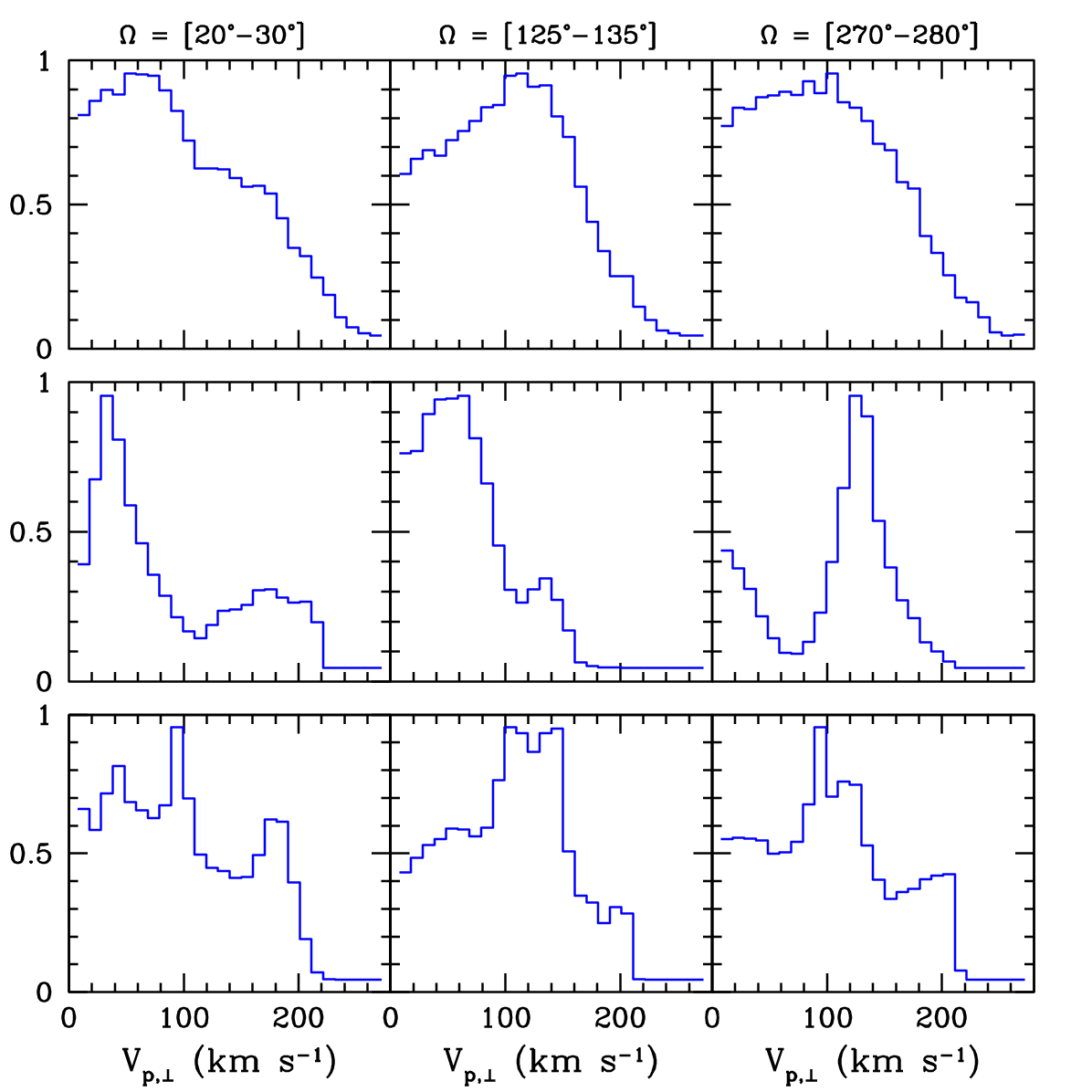}
\caption{Probability distributions for the component of the peculiar
   velocity of the binary perpendicular to the orbital plane. 
   The histograms are obtained with the orbit integrator 
   described in \S~3.1 assuming an inclination angle of the 
   orbital plane $i = 70^{\circ}.$  The distributions in the upper 
   panel are for a supernova event occurred 
   $\tau_{\rm ex}=3 \gyr$ ago in the Galactic disk and for 
   different possible ranges of values of the position angle of the 
   line of the node $\Omega$.
   The distributions in the central panel are for the same parameters
   but reducing of a factor 10 the error bars on the available 
   determination of the current 3D-velocity of the system.
   The distributions in the lower panel are as in the central panel
   but summed over all the disk crossings between 0.5 and 3 Gyr ago.}
\end{center}
\end{figure}

\section{Conclusions}
We have investigated the origin of the large peculiar velocity and
high-latitude orbit of the soft X-ray transient $\xte$. 
We constrain the origin of the binary system by studying the evolutionary
state of the companion star and by calculating the orbital trajectory
of the binary system in the potential of the Galaxy.

Based on the high CNO enrichment of the donor star, we argue that it
was born as a $1.2-1.8\msun$ main-sequence star, with a most
probable value of $1.5\msun$ (Haswell et al 2002). 
The turn-off age of such a star is between 1 and 4.5\,Gyr. 
The subsequent evolution as a semi-detached binary lasts for
about $0.8\gyr$, providing a tentative upper limit to the age of the
binary system of about 2 to 5.5\,Gyr, which is much smaller than the
typical age of globular clusters. We therefore conclude that the binary
cannot have formed in a globular cluster, but must have formed in the
Galactic disk.

In that case, the binary must have been propelled in its current
galactic high latitude orbit by the supernova explosion in which the
black hole formed. This must have happened shortly after the formation
of the binary as the progenitor of the black hole lives very short
compared to the companion star.

We calculate the orbital trajectory of the binary backward in time
through the potential of the Galaxy for a time comparable to the age 
of the system to find the average peculiar velocity of the system
at the location where it was ejected about $0.5-5\gyr$ ago. 
From these calculations we conclude that, upon birth, the binary 
system must have received an average kick of $183\pm31\kms$. 

The maximum systemic velocity which can be acquired by
the binary upon symmetric mass loss is about $100-120\kms$. 
Therefore, a symmetric supernova is not likely 
(on a 2.7\,$\sigma$ level) to have propelled the binary in its current orbit
but an additional asymmetric kick is required.
Using the available information on the 3D orientation of 
the binary,  we derive a probability of 95\% for the component of the kick
perpendicular to the orbital plane to exceed the random vertical motion 
in the Galactic disk. 
The average of this distribution is $93^{+55}_{-60}\kms$.
These results are rather insensitive to the exact age of the system
but depends heavily on the kinematic parameters of the system. 
The measurements of the spatial velocity of the binary
should be ten times more accurate than currently available 
in order to draw tight values on the asymmetric kick.

\section {Acknowledgments}
We would like to thank Jasinta Dewi for performing binary
evolutionary calculations relevant for this work,
Franca D'Antona for useful discussions and Gijs Nelemans 
and Tom Maccarone for useful comments on the manuscript.
This work was supported by the Netherlands Organization 
for Scientific Research (NWO), the Royal Netherlands Academy 
of Arts and Sciences (KNAW), the Netherlands Research School 
for Astronomy (NOVA) and the Italian Ministry of University and
Research (MIUR) under the national program {\em Cofin 2003}".

\end{document}